\documentclass[12pt,preprint]{aastex}

\slugcomment{LA-UR-03-9129}
\shorttitle{Refraction in stellar atmospheres}
\shortauthors{Kowalski and Saumon}

\def\Teff{T_{\rm eff}}
\def\sss{\scriptscriptstyle}
\def\wig#1{\mathrel{\hbox{\hbox to 0pt{%
          \lower.5ex\hbox{$\sim$}\hss}\raise.4ex\hbox{$#1$}}}}
\begin{document}

\title{Radiative transfer in the refractive atmospheres of very cool white dwarfs}
\author{P.M. Kowalski\altaffilmark{1,2} and D. Saumon\altaffilmark{2,1}}

\affil{\altaffilmark{1}Department of Physics and Astronomy, Vanderbilt University, Nashville, TN 
37235-1807}
\affil{\altaffilmark{2}Los Alamos National Laboratory, MS F699, Los Alamos, NM 87545}
\email{kowalski@lanl.gov}

\begin{abstract}

We consider the problem of radiative transfer in stellar atmospheres where the 
index of refraction departs from unity 
and is a function of density and temperature. We present modified Feautrier and 
Lambda-iteration methods to solve 
the equation of radiative transfer with refraction in a plane parallel atmosphere. 
These methods are general and can be used in any problem with 1-D geometry
where the index of refraction is a monotonically varying function of vertical optical depth.
We present an application to very cool white dwarf atmospheres where the index of refraction 
departs significantly from unity.
We investigate how ray curvature and total internal reflection affect the limb darkening 
and the pressure-temperature structure of the atmosphere. 
Refraction results in a much weakened limb darkening effect. 
We find that through the constraint of radiative equilibrium, total internal 
reflection warms the white dwarf atmosphere near the surface ($\tau \wig< 1$). This 
effect may have a significant impact on studies of very cool white dwarf stars. 
\end{abstract}


\section{Introduction}

\begin{figure*}[t]
\centering
\epsscale{0.5}
\plotone{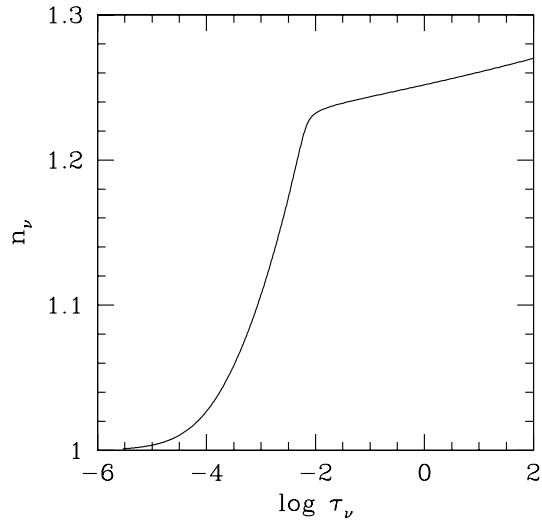}
\caption{Variation of the index of refraction as a function of vertical optical 
       depth $\tau_{\nu}$ in the nominal white dwarf atmosphere model for $\lambda=0.948\,\mu$m. 
       \label{F1}}
\end{figure*}

Problems involving radiative transfer in refractive media appear to have received limited attention 
\cite{harris65, zhelez67, pomranin68}.  In stellar atmospheres in particular,
refraction is always ignored since the gas is so tenuous that the index of refraction does not 
depart from unity. A notable exception 
is found in the atmospheres of very cool white dwarfs, especially those rich in 
helium, where the gas (or rather the fluid) density can reach
$0.1-2\,$g/cm$^3$. Under these conditions, the refractive index of fluid, atomic 
helium can become as large as $\sim 1.3$  
with very large gradients (Fig. \ref{F1}) and refraction can be expected to affect 
the radiation field. 
In this contribution, we study radiative transfer in a stellar atmosphere with 
refraction. While we are 
motivated primarily by astrophysical applications, the method developed here is 
general and can be directly 
applied to other problems with plane parallel geometry with an arbitrary, monotonic
variation of the index of refraction.

We consider refraction in a stellar atmosphere that is static, plane parallel, 
in local thermal equilibrium, in hydrostatic
equilibrium, with both radiative and convective energy transport.  The total flux is constant throughout the atmosphere and
scattering is assumed isotropic.  We further assume that the refraction of ray paths follows Snell's law (i.e. geometric optics)
as is appropriate in white dwarf atmospheres.  For a given run of the index of refraction through the atmosphere, this
completely defines the trajectories of the ray paths and allows the derivation of a simple form of the equation of radiative 
transfer (ERT) in the presence of dispersive effects.

\begin{figure*}[t]
\centering
\epsscale{0.8}
\plotone{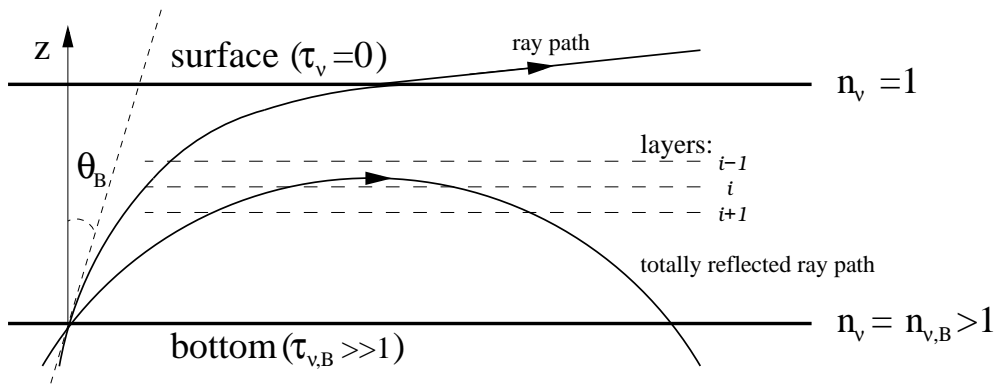}
\figcaption[P2.eps]{Geometry and typical ray paths for an axially symmetric radiation field in 
       a refractive, planar atmosphere with a monotonically varying refractive index $n_{\nu}$.
       \label{F2}}
\end{figure*}

There is a number of interesting physical effects that we can expect in a stellar 
atmosphere where the
index of refraction is greater than unity.  The index $n_\nu$ is a monotonically 
increasing function of density.
In a stellar atmosphere, the index decreases from the bottom toward the surface 
where $n_\nu=1$ at $\tau_\nu=0$.
All rays are refracted away from the upward vertical and some rays are internally 
reflected back toward the
interior (Fig. \ref{F2}).  Intuitively, this should lead to an increase in the 
temperature profile of the atmosphere to achieve flux conservation.
Because the largest variations of the index occur in the optically thin regions 
of the atmosphere (Fig. \ref{F1})
we expect refraction to be mainly a surface effect.
Furthermore, the group velocity of light is reduced by a factor of $n_\nu$ so light 
accelerates toward the
surface where it reaches the speed of light in a vacuum.  This directly affects the flux 
through its dependence on the velocity of propagation.  The angular redistribution of  light
due to refraction will reduce the limb darkening effect.
Finally, the index of refraction of fluid helium is nearly independent of frequency 
from the optical to the near 
infrared and there should not be any significant chromatic effects.

We follow the treatment of the theory of radiative transfer in dispersive media presented by 
Cox \& Giuli (1968).
We develop two different numerical schemes to solve the ERT in the presence of 
refraction based on the 
Feautrier and $\Lambda$-iteration methods. Both methods are widely used in stellar 
atmosphere codes and the
procedures we have developed allows for a straightfroward implementation of the 
effects of refraction.
To illustrate the effects of refraction in a stellar atmosphere, we apply both methods to solve for
the radiation field in a realistic helium-rich white dwarf atmosphere model 
with $\Teff=4000\,$K, a gravity of $\log g ({\rm cm/s}^2)=8$  and a homogeneous
composition of $n({\rm He})/n({\rm H})=10^6$, where $n({\rm X})$ is the number 
density of element X.  This atmospheric structure, obtained with our white dwarf
atmosphere code (Bergeron, Saumon \& Wesemael 1995), is hereafter referred to as the 
``nominal" model.  

In the next section, we construct the ERT for a refractive, planar atmosphere, and new expressions 
for the moments of the radiation field are given in Section 3. The numerical solution with
modified Feautrier and $\Lambda$-iteration methods is described in Section 4.  
Numerical aspects of the solution, a detailed discussion of the effects of refraction on the 
radiation field, and the effects on
the structure of the atmosphere and the emergent flux are presented in Section 5.

\section{The equation of radiative transfer in the presence of refractive index gradient}

The ERT along an arbitrary ray path is \cite{4}
\begin{equation}  
    \frac{dI_{\nu}(\vec r,\vec q)}{\rho ds}=
    j_{\nu}(\vec r,\vec q)-\chi_\nu(\vec r,\vec q)I_\nu(\vec r,\vec q)+
    \left({ dI_{\nu}(\vec r,\vec q)\over \rho ds}\right)_{ref} \label{5} 
\end{equation}
where $I_{\nu}$ is the specific intensity at point $\vec r$ inside the atmosphere, $\vec q$ 
is the direction of the curved ray path, $\rho$ is the mass density, $ds$ is an element 
along the curved ray path, $j_{\nu}$ is the total mass emission
coefficient, and 
\begin{equation}
    \chi_\nu=\kappa_\nu + \sigma_\nu
\end{equation}
is the total mass absorption coefficient, the
sum of true absorption ($\kappa_\nu$) and scattering ($\sigma_\nu$) processes. The mass 
emission and absorption coefficients
are related to their non-dispersive values $j_{\nu}^0$, $\chi_{\nu}^0$ by 
$j_{\nu}=n_{\nu}j_{\nu}^0$, and
$\chi_{\nu}=\chi_{\nu}^0/n_{\nu}$, respectively \cite{harris65}.  
The last term on the r.h.s. of Eq. (1) includes the contribution to
$dI_{\nu}/\rho ds$ coming from spatial variations of the refractive index. 
In a horizontally homogeneous medium, the geometry leads to an axially symmetric radiation field.
In the presence of refraction, the ray path is defined by Snell's law:
\begin{equation} n_{\nu}\sin(\theta)=n'_{\nu}\sin(\theta')=\rm constant \label{6}\end{equation}
\begin{equation} \phi=\phi' \pm \pi\end{equation}
where $\theta$ and $\phi$ are the polar angles coordinates with respect to the vertical $z$ axis, parallel
to the direction of the gradient of $n_{\nu}$.  
Applying the law of energy conservation to an incident beam of radiation with solid angle $d\omega$, 
propagating from a medium with refractive index $n_{\nu}$ 
into a medium with refractive index $n_{\nu}'$, and refracted into $d\omega'$,
\begin{equation} I_{\nu}\cos(\theta)d\omega=I'_{\nu}\cos(\theta')d\omega'\end{equation}
we find that ${I_{\nu}}/{n_{\nu}^{2}}$ is constant along a ray path if there are no energy losses or gains
due to emission, absorption, or interface effects \cite{4}. On the basis of that assumption,
the last term in equation (\ref{5}) is 
\begin{equation} \left(\frac{dI_{\nu}}{ds}\right)_{ref}=\frac{\partial I_{\nu}}{\partial n_{\nu}}
\frac{d n_{\nu}}{ds}=\frac{2I_{\nu}}{n_{\nu}}\frac{d n_{\nu}}{ds}\end{equation}
and the ERT becomes
\begin{equation}  \frac{d}{\rho ds}\left(\frac{I_{\nu}(\vec r,\vec q)}{n_{\nu}^{2}}\right)=
\frac{j_\nu(\vec r,\vec q)-\chi_\nu(\vec r,\vec q)I_{\nu}(\vec r,\vec q)}
{n_{\nu}^{2}}. \label{12}\end{equation}

In plane-parallel geometry, the derivative $d/ds$ takes the form
\begin{equation} \frac{d}{ds}=\cos\theta {\partial \over \partial z} +
{d \theta \over ds}{\partial \over \partial \theta}. \label{13}\end{equation}
The derivative over $\theta$ appears because the ray paths are curved. 
For a given ray path, however, Eq. (\ref{6}) 
allows us to reduce the configuration space from $\{z,\theta\}$ to a one-dimensional curve
in $\{z\}$ only. 
The ray paths are parameterized by $\theta_B$, the value of the ray path angle at a reference level
chosen to be the bottom of the atmosphere (Fig. \ref{F2}).
Hereafter, all quantities with subscript $``B"$ refer to this lower boundary. A ray path is described by
\begin{equation} \mu_{B}=\cos\theta_{B}={\rm constant} \label{14}\end{equation}
and as for each one dimensional curve $dz/ds=\cos\theta$, we can write
\begin{equation} \frac{d}{ds}=\cos\theta{\partial \over \partial z}\end{equation}
and rewrite the ERT (\ref{12}) in a simpler form
\begin{equation} {\partial I'_{\nu}(\tau_{\nu},\mu_{B})\over\partial \sigma_{\nu}}
=I'_{\nu}(\tau_{\nu},\mu_{B})-S'_{\nu}(\tau_{\nu})  \label{17} \end{equation}
or
\begin{equation} 
    \mu(\mu_{B},n(\tau_{\nu})) {\partial I'_{\nu}(\tau_{\nu},\mu_{B})\over\partial \tau_{\nu}}
    =I'_{\nu}(\tau_{\nu},\mu_{B})-S'_{\nu}(\tau_{\nu}),  \label{18} 
\end{equation}
where a primed quantity $f'$ represents $f/n_{\nu}^{2}$, $S_{\nu}=j_{\nu}/\chi_{\nu}$ is the 
source function, $\sigma_{\nu}$ is now the optical depth
measured along the curved ray path, $\tau_{\nu}$ is the vertical optical depth, and  
\begin{equation} 
    \mu={\rm sign} (\mu_{B})\sqrt{1-\left(\frac{n_{\nu,B}}{n_{\nu}}\right)^{2}(1-\mu_{B}^{2})} \label{19}.
\end{equation}
The optical depths $\sigma_{\nu}$ and $\tau_{\nu}$ are related 
to physical distances through the relations: 
$d\tau_{\nu}=-\rho\chi_{\nu}dz$ and $d\sigma_{\nu}=-\rho\chi_{\nu}ds$.

Equation (\ref{18}) reduces to the well-known expression for the ERT for a non-refractive, plane-parallel atmosphere
if we set $\mu=\rm constant$ and $n_{\nu}=1$.
In the refractive case, however, the angle $\mu$ is now a function of $\mu_{B}$ and $\tau_{\nu}$, and
at any given level, only the rays that started at the bottom with $\mu_B$ given by
\begin{equation} \mu_{Bmin}^2(\tau_\nu) = 1-\left({n_\nu(\tau_\nu) \over n_{\nu,B}}\right)^2<\mu_B^2<1.\end{equation}
will not have been reflected downward.
We label the set of all possible values of $\mu_{B}$ at a given $\tau_\nu$  by $\zeta$.

While Eqs. (\ref{17}) and (\ref{18}) are mathematically equivalent forms of the ERT, 
their numerical solution with finite difference schemes have very different accuracies for strongly curved ray paths.
Equation (\ref{18}) is simpler to solve numerically, as it does not require the calculation of $\sigma_{\nu}$, however
we find that it is essential to integrate the optical depth along the curved ray path to obtain a good solution 
and we will use the form given by Eq. (\ref{17}).   
 
\section{Moments of the radiation field}

The moments of the radiation field $J_{\nu}(\tau_{\nu})$,
$H_{\nu}(\tau_{\nu})$, and $K_{\nu}(\tau_{\nu})$ are defined as
\begin{equation} J_{\nu}(\tau_{\nu})=\frac{1}{2}\int\limits_{-1}^{1}{I_{\nu}(\tau_{\nu},\mu)\,d\mu},\end{equation}
\begin{equation} H_{\nu}(\tau_{\nu})=\frac{1}{2}\int\limits_{-1}^{1}{\mu I_{\nu}(\tau_{\nu},\mu)\,d\mu}, \label{321}\end{equation}
and
\begin{equation} K_{\nu}(\tau_{\nu})=\frac{1}{2}\int\limits_{-1}^{1}{\mu^{2} I_{\nu}(\tau_{\nu},\mu)\,d\mu}.\end{equation}
The ERT for the non-refractive case can be rewritten in terms of these moments:
\begin{equation} \frac{\partial H_{\nu}(\tau_{\nu})}{\partial \tau_{\nu}}=J_{\nu}(\tau_{\nu})-S_{\nu}(\tau_{\nu}), \label{24}\end{equation}
\begin{equation} \frac{\partial K_{\nu}(\tau_{\nu})}{\partial \tau_{\nu}}=H_{\nu}(\tau_{\nu}) \label{25},\end{equation}
and
\begin{equation} \frac{\partial^{2} K_{\nu}(\tau_{\nu})}{\partial \tau_{\nu}^{2}}=J_{\nu}(\tau_{\nu})-S_{\nu}(\tau_{\nu}). \label{26}\end{equation}
For the more general refractive case, the moments of the radiation field become  
\begin{equation} J'_{\nu}(\tau_{\nu})=\frac{1}{2}\int\limits_{-1}^{1}{I'_\nu(\tau_\nu,\mu)\,d\mu}
=\frac{1}{2}\int\limits_{\zeta(\tau_{\nu})}{I'_{\nu}(\tau_{\nu},\mu_B){\partial \mu \over \partial \mu_B}\,d\mu_B},
\label{27}\end{equation}
\begin{equation} H'_{\nu}(\tau_{\nu})=\frac{1}{2}\int\limits_{-1}^{1}{\mu I'_{\nu}(\tau_{\nu},\mu)\,d\mu} 
=\frac{1}{2}\int\limits_{\zeta(\tau_\nu)}{\mu(\mu_B) I'_\nu(\tau_\nu,\mu_B){\partial \mu \over \partial \mu_B}\,d\mu_B},
\label{28}\end{equation}
and
\begin{equation} 
    K'_{\nu}(\tau_{\nu})=\frac{1}{2}\int\limits_{-1}^{1}{\mu^{2} I'_{\nu}(\tau_{\nu},\mu)\,d\mu}
    =\frac{1}{2}\int\limits_{\zeta(\tau_\nu)}{\mu^2(\mu_B)I'_\nu(\tau_\nu,\mu_B){\partial \mu \over \partial \mu_B}\,d\mu_B},
    \label{29}
\end{equation}
where
\begin{equation} {\partial \mu \over \partial \mu_{B}}=\frac{\eta (\tau_{\nu})}{\mu},\label{30}\end{equation}
and
\begin{equation} \eta (\tau_{\nu})=\mu_{B}\left(\frac{n_{v,B}}{n_{\nu}}\right)^{2}. \label{31}\end{equation}
In the refractive case, $J^{'}_{\nu}$, $H^{'}_{\nu}$, and $K^{'}_{\nu}$ are more convenient quantities
than $J_{\nu}$, $H_{\nu}$ and $K_{\nu}$ since $I^{'}_{\nu}$ is constant along a ray path.
The construction of the moments equations (\ref{24})--(\ref{26}) is complicated by the 
fact that $\mu$ and the integration domain $\zeta$ are now functions of $\tau_\nu$.
We first consider the derivative of $H'_{\nu}$ over $\tau_{\nu}$.
For $\mu$ and $\eta$ given by equations (\ref{19}) and (\ref{31}) we get
\begin{equation} 
    {\partial \eta \over \partial \tau_\nu}=-2\eta\frac{1}{n_\nu}\frac{\partial n_\nu}{\partial \tau_\nu}
\end{equation}
and
\begin{equation} 
    \frac{\partial \mu}{\partial \tau_\nu}=-\frac{1-\mu^2}{\mu}\frac{1}{n_\nu}\frac{\partial n_\nu}{\partial \tau_\nu}. \label{311} 
\end{equation}
Taking the derivative of (\ref{28}) over $\tau_{\nu}$ we obtain
\begin{eqnarray}
    \frac{\partial H_{\nu}^{'}}{\partial \tau_{\nu}} & = & 
    \frac{1}{2}\frac{\partial}{{\partial \tau_{\nu}}} \int_\zeta{\eta I^{'}_\nu \,d\mu_{B}} 
    = \frac{1}{2}\int_{\zeta}{\frac{\partial}{\partial \tau_{\nu}}\left(\eta I_{\nu}^{'}\right)\,d\mu_B} \nonumber \\ 
    & & \quad\quad -  \frac{\partial \mu_{Bmin}}{{\partial \tau_{\nu}}}\eta I^{'}_{\nu}\Big|_{\mu_{B} = \mu_{Bmin}}+
    \frac{\partial (-\mu_{Bmin})}{\partial \tau_{\nu}}\eta I^{'}_{\nu}\Big|_{\mu_{B}=-\mu_{Bmin}}. 
\end{eqnarray}
Since $I_{\nu}^{'}(\mu_{Bmin})=I_{\nu}^{'}(-\mu_{Bmin})$ and $\eta(\mu_{Bmin})=-\eta(-\mu_{Bmin})$, we get
\begin{eqnarray} 
    \frac{\partial H_{\nu}^{'}}{\partial \tau_\nu} & = & 
    \frac{1}{2}\int_\zeta{\frac{\partial}{{\partial \tau_\nu}}(\eta I_\nu^{'})\,d\mu_B} \nonumber \\
    & = & \frac{1}{2}\int_\zeta{\frac{\partial \eta}{\partial \tau_\nu}(I_\nu^{'})\,d\mu_B} + \frac{1}{2} 
    \int_{\zeta} {\eta \mu^{-1}(I'_\nu- S'_\nu)\,d\mu_B} \nonumber \\ 
    & = & J'_\nu- S'_\nu-2H'_\nu\frac{1}{n_\nu}\frac{\partial n_\nu}{\partial \tau_\nu},
\end{eqnarray}
and finally
\begin{equation} \frac{\partial H_{\nu}(\tau_{\nu})}{\partial \tau_{\nu}}=
J_{\nu}(\tau_{\nu})-S_{\nu}(\tau_{\nu}), \end{equation}
which is identical to equation (\ref{24}). The condition for radiative equilibrium
\begin{equation} 
    {\partial \over \partial z} \int_0^\infty {H_{\nu}\,d\nu}=\int_0^\infty \chi_{\nu}(J_{\nu}-S_{\nu})\,d\nu=0 \label{37}
\end{equation}
remains unchanged.  An analogous derivation for $\partial K'_{\nu} / \partial \tau_{\nu}$ gives
\begin{equation} 
    \frac{\partial K_{\nu}^{'}}{\partial \tau_{\nu}} = H'_{\nu}(\tau_{\nu}) + 
    \frac{1}{2}\int_{\zeta}{\frac{\partial (\mu\eta)}{\partial \tau_{\nu}} I^{'}_{\nu}\,d\mu_B} - 
    \frac{1}{2}\int_{\zeta}{\eta S^{'}_{\nu}\,d\mu_{B}}. \label{39}
\end{equation}
Since $\eta$ is antisymmetric in $\mu_{B}$, the last term in equation (\ref{39}) is $0$ and we finally get
\begin{equation} 
     \frac{\partial K_{\nu}^{'}}{\partial \tau_{\nu}} = H'_{\nu}(\tau_{\nu}) -
     (3K'_{\nu}-J'_{\nu})\frac{1}{n_{\nu}}\frac{\partial n_{\nu}}{\partial \tau_{\nu}}
\end{equation}
or
\begin{equation} 
     \frac{\partial K_{\nu}}{\partial \tau_{\nu}} = H_{\nu}(\tau_{\nu}) -
     (K_{\nu}-J_{\nu})\frac{1}{n_{\nu}}\frac{\partial n_{\nu}}{\partial \tau_{\nu}} \label{41}.
\end{equation}
Equation (\ref{41}) contains an additional term due to refraction. In cool white dwarf atmospheres, this term can 
dominate near the surface, as the gradient of the index of refraction becomes very large (Fig. \ref{F1}).
This equation may be used to evaluate the radiative flux $H_{\nu}$ when $J_{\nu}$, $K_{\nu}$, and $n_{\nu}$ are known.
The form of Eq. (\ref{26}) that includes refraction contains a term in $\partial J_\nu/\partial \tau_\nu$ and is not useful.

\section{Solution of the equation of radiative transfer in a dispersive medium}
\subsection{Feautrier solution}

In the presence of refraction, it remains advantageous to define the symmetric average of the specific intensity
\begin{equation} 
    P'(\mu_{B},v,\sigma_{\nu})=\frac{1}{2}[I'(\mu_{B},\nu,\sigma_{\nu})+I'(-\mu_{B},\nu,\sigma_{\nu})]. \label{45}
\end{equation}
Differentiating with respect to $\sigma_{\nu}$, we obtain
\begin{equation} 
    {\partial P'(\mu_{B},\nu,\sigma_{\nu}) \over \partial \sigma_{\nu}}=\frac{1}{2}[I'(\mu_{B},\nu,\sigma_{\nu})
    -I'(-\mu_{B},v,\sigma_{\nu})] \label{46} 
\end{equation}
and
\begin{equation} 
    {\partial^{2}P'(\mu_{B},\nu,\sigma_{\nu}) \over \partial \sigma_{\nu}^{2}}
    =P'(\mu_{B},\nu,\sigma_{\nu})-S'(\sigma_{\nu}). \label{47}
\end{equation}
The source function in a dispersive medium and under the assumption of LTE and isotropic scattering is \cite{4} 
\begin{equation} 
    S'_{\nu}(\tau_{\nu})=\epsilon_{\nu}B_{\nu}+(1-\epsilon_\nu)J'_{\nu}, \label{SF1} 
\end{equation}
where $\epsilon_\nu=\kappa_\nu/\chi_\nu$ is an absorption coefficient (equivalently, $1-\epsilon_\nu$ is
the scattering coefficient, or albedo). 
Equations (\ref{45}--\ref{47}) are mathematically identical to those used of the non-refractive case 
(Mihalas 1978), and their solution will require only minor changes. 

Two boundary conditions are required to 
solve the second-order differential equation (\ref{47}).  
At the bottom of the atmosphere, where $\tau_{\nu}>>1$, the radiation field is very close to thermodynamical equilibrium and
\begin{equation} P'_{B}=B_{\nu} \label{48}\end{equation}
where $B_{\nu}$ is the Planck function. 
To ease the notation, we drop the subscript $\nu$ in the remainder of our discussion of the solution
of the ERT (Eqs. 39 -- 51).

We consider that there is no incident radiation on the top of the atmosphere ($\tau_{\nu}=0$). For ray paths 
that exit at the surface ($\mu_{B}>\mu_{Bmin}(0)$), the surface boundary condition is obtained from Eqs. (\ref{45}) and (\ref{46}):
\begin{equation} 
    {\partial P' \over \partial\sigma}\Bigg|_{\sigma_{surface}}=P'_{\sss 0}.
\end{equation}
Because of refraction, rays with $\mu_{B}<\mu_{Bmin}(0)$ are reflected downward (Fig. \ref{F2}). These ray paths 
require a different boundary condition. At the reflection point, $\mu=0$, $I'(+\mu)=I'(-\mu)$, and we have (\ref{46})
\begin{equation} 
    {\partial P' \over \partial\sigma}\Bigg|_{\sigma_{reflection \ point}}=0.
\end{equation}
We solve the equations in finite difference form, where the vertical structure of the atmosphere is divided in 
$i=1,$ 2, ... $,N$ horizontal layers with $i=1$ corresponding to the topmost layer.  
We develop higher-order expressions for the boundary conditions using Taylor expansions:
\begin{equation} 
    P'_{i+1}=P'_{i}+{\partial P' \over \partial \sigma}\Bigg|_{\sigma_{i}}\Delta\sigma_{i+1}+{1 \over 2} 
    (P'_{i}-S'_{i}) \Delta\sigma_{i+1}^{2}
\end{equation}    
\begin{equation} 
    {\partial P' \over \partial \sigma}\Bigg|_{\sigma_{top}}={\partial P' \over \partial \sigma}\Bigg|_{\sigma_{i}}+
    (P'_{i}-S'_{i})(\sigma_{top}-\sigma_{i})
\end{equation}
where $\Delta\sigma_{i}=\sigma_{i}-\sigma_{i-1}$ and $\sigma_{top}$ is the optical depth at the  boundary (either the surface or 
the reflection point) measured along the ray path.  For the surface boundary condition we get
\begin{equation} 
    {{P'_{2}-P'_{1}} \over \Delta\sigma_{2}}=P'_{1}+{1 \over 2} (P'_{1}-S'_{1})(\sigma_{1}+\sigma_{2}-2\sigma_{top})\label{53}
\end{equation}
and for the boundary condition for ray paths that are totally reflected between levels $\tau_{R}$ and $\tau_{R-1}$
\begin{equation} {{P'_{R+1}-P'_{R}} \over \Delta\sigma_{R+1}}={1 \over 2} (P'_{R}-S'_{R})(\sigma_{R}+\sigma_{R+1}-2\sigma_{top}).\label{54}\end{equation}

Equations (\ref{47}), (\ref{48}), (\ref{53}), and (\ref{54}) are solved by modifying the Feautrier solution \cite{8}. 
The derivatives of any physical quantity $f$ with respect to $\sigma$ are evaluated with finite differences:
\begin{equation} 
    \left({df \over d\sigma}\right)_{i-{1 \over 2}} = {{f_{i}-f_{i-1}} \over {\sigma_{i}-\sigma_{i-1}}},
\end{equation}      
and for the second derivative
\begin{equation} 
    \left({d^{2}f \over d\sigma^{2}}\right)_{i}={{\left({df \over d\sigma}\right)_{i+{1 \over 2}}- 
    \left({df \over d\sigma}\right)_{i-{1 \over 2}}} \over {{1 \over 2} (\sigma_{i+1}-\sigma_{i-1})}}.
\end{equation}
The ERT is solved on a discrete grid of $\mu_{B}$ points, that defines the ray paths. 
Each ray path is labeled $\mu_{B,k}$, where $k=1,...,D$. 
The set of radiative transfer equations (\ref{47}), can then be written as an algebraic matrix equation \cite{8}:
\begin{equation} 
    \bf- A_{\it i}\bf P'_{\it i \rm -1}+ \bf B_{\it i} \bf P'_{\it i}- \bf C_{\it i} \bf P'_{\it i \rm +1}= \bf L_{\it i},
\end{equation}
where $\bf A_{\it i}$, $\bf C_{\it i}$, and $\bf L_{\it i}$ are $D \times 1$ vectors corresponding to 
layer $\it i$, with row $\{\bf A_{\it i}\it\}_{j}$,$\{\bf C_{\it i}\it\}_{j}$,$\{\bf L_{\it i}\it\}_{j}$ 
corresponding to ray path $j$, and $\bf B_{\it i}$ is a 
$D \times D$ matrix, corresponding to layer $\it i$ with elements $\{\bf B_{\it i} \it \}_{j,k=1,\ldots,D}$ 
corresponding to ray path $j$. 
The surface and bottom boundary conditions are constructed from equations (\ref{53}) and (\ref{48}). 

In the presence of refraction we need to consider the treatment of totally reflected ray paths. 
According to Eq. (\ref{54}), for ray path $j$ reflected between layers $R$ and $R-1$, the boundary 
condition for reflected rays can be written as 
\begin{equation} 
    \sum_{k} \{\bf B_{\it R} \it \}_{j,k}\{\bf P'_{\it R}\it \}_{k}-\{\bf C_{\it R}\it \}_{j}\{\bf P'_{\it R \rm +1}\it\}_{j}=
    \{\bf L_{\it R} \it \}_{j}. 
\end{equation}
Furthermore, the number of rays in a given layer decreases toward the top of the atmosphere. To 
maintain the dimensions of $\bf B_{\it i}$, the elements
corresponding to reflected ray paths are treated differently.  
For angles $\mu_{B,j}<\mu_{Bmin}(\tau_{i})$, i.e. for ray paths reflected deeper than $\tau_{i}$, we set:
\begin{equation} \{\bf B_{i} \it\}_{j,k}=\cases{0, &for $ j \ne k $ \cr 1, &for $ j = k$}\end{equation}
and
\begin{equation} \{\bf A_{i} \it \}_{j}=\{\bf C_{i} \it \}_{j}=\{\bf L_{i} \it \}_{j}=\rm 0.\end{equation}

The solution of Eq. (47) for $P'(\tau_\nu,\mu_B)$ is otherwise identical to the non-refractive case \cite{8}:
\begin{eqnarray} 
    \bf P'_{\it i} & = & \bf D_{\it i} \bf P'_{\it i \rm +1}+\bf v_{\it i} 
    \nonumber \\ \bf D_{\it i} & = & \bf (B_{\it i \bf}-A_{\it i \bf }B_{ \it i \rm -1})^{-1} \bf C_{\it i}
    \nonumber\\ \bf v_{\it i} & = & \bf (B_{\it i \bf }-A_{\it i \bf }B_{ \it i \rm -1})^{-1}
                ( \bf L_{ \it i \bf }+A_{ \it i} \bf v_{ \it i \rm -1})
    \nonumber\\ \bf D_{\rm 1} & = & \bf B_{\rm 1}^{-1} \bf C_{\rm 1}
    \nonumber\\ \bf v_{\rm 1} & = & \bf B_{\rm 1}^{-1} \bf L_{\rm 1}, 
\end{eqnarray}
and the moments of the radiation field are then obtained with
\begin{equation} 
    J'_{\nu}(\tau_{\nu})= \frac{1}{2}\int\limits_{0}^{1} {P'_{\nu}(\tau_{\nu},\mu_{B})\,d\mu},
\end{equation}
\begin{equation} H'_{\nu}(\tau_\nu)=
    \frac{1}{2}\int\limits_{0}^{1} {\mu^2{\partial P'_\nu(\tau_\nu,\mu_B) \over \partial \tau_{\nu}}\,d\mu}
    =\frac{1}{2}\int\limits_{0}^{1} {\mu {\partial P'_\nu(\sigma_\nu,\mu_B) \over \partial \sigma_\nu}\,d\mu},
    \label{55}
\end{equation}
and
\begin{equation} 
    K'_{\nu}(\tau_{\nu})= \frac{1}{2}\int\limits_{0}^{1} {\mu^2 P'_\nu(\tau_\nu,\mu_B)\,d\mu}.
\end{equation}

\subsection{$\Lambda$-iteration method}

Given the formal solution of the ERT along a ray path between optical depths 
$\sigma_{\nu,1}$ and $\sigma_{\nu,2}$ ($\sigma_{\nu,1}<\sigma_{\nu,2}$)
\begin{equation} 
    I_{\nu}'(\sigma_{\nu,1})=I_{\nu}'(\sigma_{\nu,2})e^{-(\sigma_{\nu,2}-\sigma_{\nu,1})} +
    \int\limits_{\sigma_{\nu,1}}^{\sigma_{\nu,2}}{S'_{\nu}(t)e^{-(t-\sigma_{\nu,1})}\,dt}, \label{67}
\end{equation}
the mean intensity (\ref{27}) is given by
\begin{equation} 
    J_{\nu}'(\tau_{\nu})=\frac{1}{2}\int_{-1}^{1}{I'_{\nu}(\sigma_{\nu}(\tau_{\nu}))\,d\mu}=\Lambda[S_{\nu}'(t)], \label{68}
\end{equation}
where $\Lambda$ is an operator which acts on the source function $S'_{\nu}$.  The source function (\ref{SF1}) is
\begin{equation} 
    S'_{\nu}(\tau_{\nu})=\epsilon_{\nu} B_{\nu}+(1-\epsilon_{\nu})\Lambda[S'_{\nu}(t)]. \label{69}
\end{equation}
An iterative method can be used to find the value of the source function $S'_{\nu}$ at a given atmosphere level
\begin{equation} 
    S'_{l+1}=\epsilon_\nu B_{\nu} +(1-\epsilon_{\nu})\Lambda S'_{l}. \label{72}
\end{equation}
This solution is known as the $\Lambda$-iteration method \cite{8}.
For a discrete grid of $\tau_{\nu}$ points, $\Lambda$ may also be represented as a matrix operator $\bf \Lambda$ that 
acts on the source function vector $\bf S'$ \cite{10}:
\begin{equation} \bf J'=\Lambda S'.\end{equation}

The elements of the matrix $\bf \Lambda$ depend only on the numerical scheme adopted for integrating over $\sigma_{\nu}$ and $\mu$ 
to evaluate the specific intensity $I'_{\nu}$, and the mean intensity $J'_{\nu}$, respectively. A good choice is to locally interpolate 
the source function by a quadratic expression in $\sigma_{\nu}$, which allows for analytic integration in (\ref{67}) and excellent numerical 
precision. In this case the construction of the $\bf \Lambda$ matrix is well-known 
and presented in details by Olson \& Kunasz (1987).  Once the source function $S'_{\nu}$ is obtained 
by solving (\ref{72}), $I'_{\nu}$ is computed from (\ref{67}), and the moments 
$J'_{\nu}$, $H'_{\nu}$ and $K'_{\nu}$ follow by direct integration. For the $\Lambda$-iteration procedure, the boundary conditions
are physically the same as in the non-refractive case: 1) $I'_{\nu}=B_{\nu}$ at the bottom of the atmosphere and 
2) there is no incident radiation at the surface $I'_{\nu}(\mu<0)=0$. There is no need for a ``surface" 
boundary condition for the totally reflected ray paths.
Since the $\Lambda$-iteration method solves directly for $I'_\nu$, we can integrate along the full length
of the reflected ray path (Eq. \ref{67}), including the downward part.    
 
We found that the $\Lambda$-iteration method works very well for cool white dwarf atmospheres. However,
it is well known that the $\Lambda$-iteration method converges poorly for nearly pure scattering media ($\epsilon_{\nu} <<1$) 
\cite{8}. This difficulty can be circumvented with the accelerated $\Lambda$-iteration methods (ALI) 
\citep{3,9,5,6}. Since the introduction of refraction does not affect the $\Lambda$-iteration method 
proper, the solution can be implemented just as easily with ALI as with the $\Lambda$-iteration.  

\section{Application to stellar atmospheres}

\subsection{Numerical considerations}

As is usual, the atmosphere is divided vertically into discrete layers, $\tau_{\nu,i}$ (Fig. \ref{F2}) and the ERT is solved for a 
finite set of ray paths $\mu_{B,k}$. Since in the refractive case the rays are no longer straight, the optical depth 
along the ray path $\sigma_{\nu}$ must be obtained by integration and it is related to the vertical optical depth $\tau_{\nu}$ by  
\begin{equation} 
\sigma_{\nu}(\tau_{\nu})=\int\limits_{\tau_{0,\nu}}^{\tau_{\nu}}{\frac{d\tau_{\nu}^{'}}{\mu(\tau^{'}_{
\nu})}} \label{75}\end{equation}
where $\tau_{0,\nu}$ is the optical depth at the bottom of the atmosphere for upward rays and $\tau_{0,\nu}=0$ 
for downward rays that start from the surface. 
We found that it is essential to calculate $\sigma_{\nu}$ precisely along the curved ray paths.
This is especially important in the layer where a given ray path is internally reflected. 
This is where the curvature of the ray is maximal and 
the ray can travel over a large horizontal distance (in optical depth) within that layer 
(Fig. \ref{F2}).  We locally interpolate 
$\mu(\tau_\nu)$ between $\tau_{\nu}$ grid points with a quadratic polynomial to 
evaluate $\sigma_{\nu}$ 
analytically. To find the vertical optical depth $\tau_\nu$ 
of the reflection point ($\mu(\tau_\nu)=0$), we use Eq. (\ref{19}) and a quadratic 
interpolation of $n_{\nu}(\tau_\nu)$.  The integrals for the moments of the radiation field are 
performed numerically with a quadratic (3-point) integration scheme. 
The integration consists of two parts: $0\le\mu\le\mu_{c}$ and $\mu_{c}\le\mu\le1$,
because $P'_\nu(\tau_\nu,\mu)$ is discontinuous at 
$\mu=\mu_{\rm c}(\tau_\nu)=\sqrt{1-n_{\nu}^{2}(\tau_{\nu}=0)/n_{\nu}^{2}(\tau_{\nu})}$ 
(see \S5.2 below).  
Rays with $\mu<\mu_{\rm c}(\tau_\nu)$ will be reflected before they reach the surface.
In Eq. (\ref{67}), $S'_\nu(\tau_\nu)$ is approximated by a quadratic polynomial in each 
layer \cite{10}.
The radiative flux from the Feautrier solution is calculated with both Eqs.  
(\ref{55}) and (\ref{41}) to check the numerical consistency of the solution.  

\begin{figure}[t]
\centering
\epsscale{0.5}
\plotone{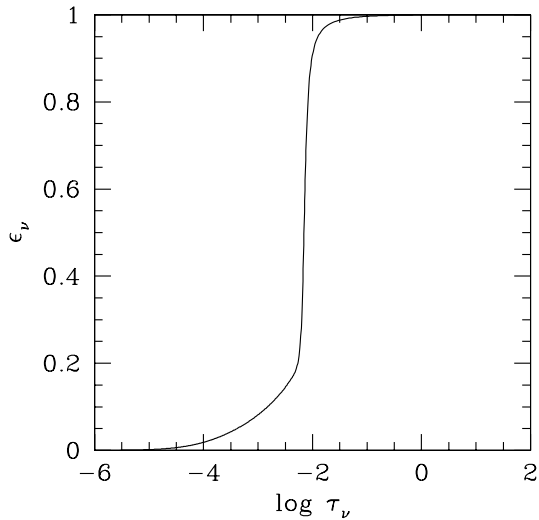}
\figcaption{Absorption parameter $\epsilon_\nu=\kappa_\nu/\chi_\nu$ in the nominal white 
       dwarf atmosphere model for $\lambda=0.948\,\mu$m. Pure scattering corresponds to $\epsilon_\nu=0$.
       \label{F3}}
\end{figure}

We have verified that our methods of solutions and codes are correct in two ways. 
To check the consistency of the solutions obtained with the Feautrier and $\Lambda$-iteration 
methods, and their sensitivity to the resolution 
of the $\tau_{\nu}$ and $\mu_B$ grids, we solved for the radiation field in a white dwarf 
atmosphere with a fixed $(T,P)$ structure. This nominal structure corresponds to a non-refractive 
helium-rich model with $\Teff=4000\,$K, $\log g=8$, $n({\rm He})/n({\rm H)}=10^{6}$ in 
radiative/convective equilibrium. In this model the convection zone extends from 
the bottom of the atmosphere up to $\tau_{\sss R} \sim 10^{-2}$.
The input parameters for computing the radiation field are
the refractive index $n_{\nu}(\tau_{\nu})$ (Fig. \ref{F1}), the absorption parameter 
$\epsilon_{\nu}(\tau_{\nu})$ (Fig \ref{F3}.), and the Planck function $B_{\nu}(\tau_{\nu})$ 
(Fig \ref{F7}.), all computed from the $(T,P)$ structure. 

For our most precise calculation, using $N=500$ layers and $D=500$ ray-paths, the relative 
differences between the two methods 
are smaller than 0.1\% in both $J_{\nu}$ and $H_{\nu}$, and the internal precision of both methods 
is better than 0.01\%. 
The difference between the two solutions decreases significantly as the number of layers increases.
With 50 layers, the moments of the radiation field can differ by as much as 1\%.   
On the other hand, the solution is much less sensitive to the number of ray paths.
In situations where the optical thickness of a layer $\tau_{\nu,i+1}-\tau_{\nu,i}\wig>1$, 
as is common in the deeper region of atmosphere models, the application 
of a quadratic interpolation of the source function $S'_\nu$ in Eq. (\ref{67}) is more 
precise than
the solution of Eq. (\ref{47}) using a finite difference scheme. The $\Lambda$-iteration 
method is therefore
a better choice when computing a model with a relatively small number of layers ($N \sim 50)$.

We have also compared with published solutions to a similar but simpler problem, where radiative 
transfer with refraction is considered in a dielectric slab \cite{1,7}. Both surfaces of the 
slab are maintained at fixed but different temperatures, radiative equilibrium is imposed, 
the index of refraction varies linearly between the two surfaces, and the 
absorption coefficient is constant. We note that Abdallah \& Dez (2000) neglected the effect 
of dispersion on the absorption coefficient, assuming that $\kappa_\nu=\kappa_\nu^0$, 
while the correct expression is $\kappa_\nu=\kappa_{\nu}^0/n_\nu$,
where $\kappa^0_\nu$ is mass absorption coefficient in the absence of 
dispersive effects \cite{harris65}. Setting $\kappa_{\nu}=\kappa_{\nu}^0$, we 
reproduce Figs. (3a--3c) 
of Abdallah \& Dez (2000) perfectly. Introducing the effect of dispersion
on $\kappa_{\nu}$ raises the temperature profile in the slab by a few degrees.  

\subsection{Refraction inside white dwarf atmospheres}

\begin{figure}[t]
\centering
\epsscale{0.75}
\plotone{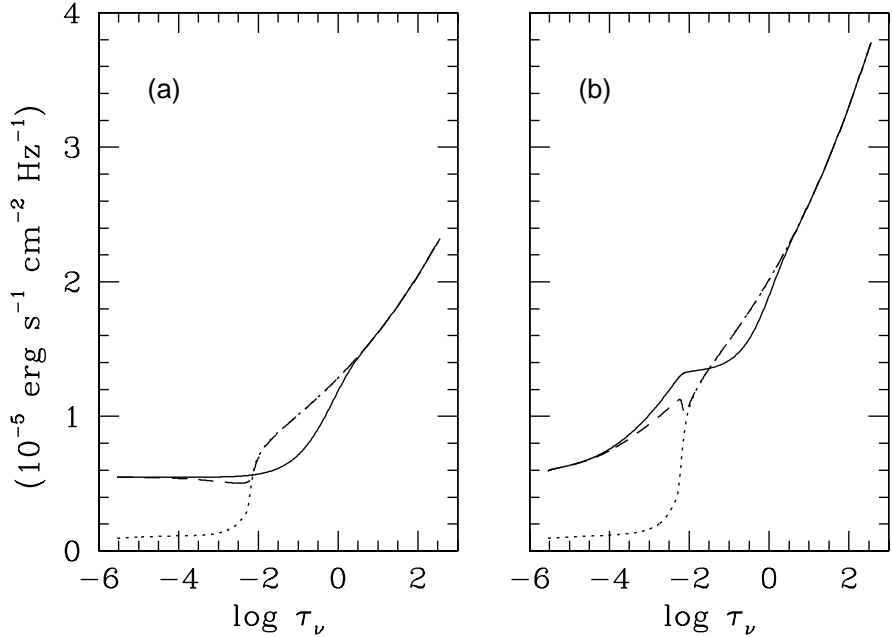}
\figcaption[P4.ps] {Mean intensity $J_{\nu}$ (solid line), source function $S_{\nu}$ (dashed line), and 
       Planck function $n_{\nu}^{2}B_{\nu}$ (dotted line) for non-refractive $(a)$ and refractive 
       $(b)$ cases as a function of vertical optical 
       depth $\tau_{\nu}$ in the nominal white dwarf atmosphere model for $\lambda=0.948\,\mu$m.
       \label{F7}}
\end{figure}
We now turn to a detailed discussion of the effects of refraction in our nominal white dwarf atmosphere model.
Figure \ref{F7} shows the mean intensity, the source function and the Planck function 
for the refractive and non-refractive cases for the same nominal $(T,P)$ structure.. 
In the refractive case, $J_{\nu}$ increases by a factor of $n^2_\nu$ at large optical depth
because $J_\nu \sim S_\nu \sim n^2_\nu B_\nu$ (Eq. \ref{SF1}). 
For $\tau_\nu<10^{-2}$, $n_\nu \sim 1$ and $n^2_\nu B_\nu \sim B_\nu$.
However, $J_{\nu}$ remains 
significantly larger than in the non-refractive case because of total internal reflection. 
This can be understood from the angular distribution of the radiation.
Figure \ref{F8} shows the angular distribution of the symmetric average of the 
specific intensity $P'_\nu$ for both cases at different levels in the atmosphere. 
Deep inside the atmosphere $P'_\nu$ is just the Planck 
function $B_\nu$ and it is the same in both cases. Toward the surface, $P'_{\nu}$ splits 
into two regions separated by a discontinuity at $\mu=\mu_{c}(\tau_\nu)$.  
The additional intensity for $\mu<\mu_{\rm c}$ arises from the contribution of 
ray paths that have been reflected at lower $\tau_{\nu}$ (above the level considered).
Toward the surface, $P'_{\nu}$ doubles across the discontinuity 
\begin{equation} 
    P'_{\nu}(\mu_{\rm c}^-) \sim 2P'_{\nu}(\mu_{\rm c}^+) 
\end{equation}
because the contribution 
from the source function from the optically thin regions lying above for $\mu>\mu_{\rm c}$ 
is negligible 
but the contribution from the downward rays ($\mu<0$) that have been reflected 
higher up is nearly the same as for the upward rays ($\mu>0$),
due to negligible absorption-emission effects inside the optically thin region. This effect 
should result in warming of an atmosphere in radiative equilibrium, because an increase 
in the mean intensity $J_{\nu}$ must generally be compensated by an increase in 
the Planck function $B_{\nu}$. 
\begin{figure}[t]
\centering
\epsscale{0.8}
\plotone{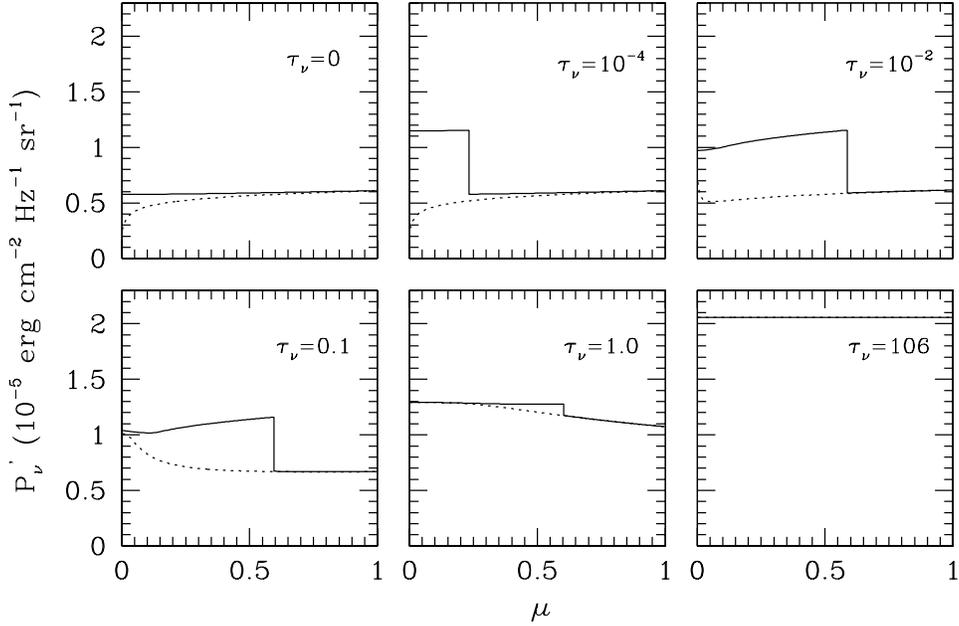}
\figcaption[P5.ps]{Symmetric average of the specific intensity 
       $P'_{\nu}=P_{\nu}/n_{\nu}^{2}$ for refractive (solid line) and non-refractive 
       (dotted line) cases as a function of angle $\mu=\cos\theta$ at various levels 
       in the nominal white dwarf atmosphere model. $\mu=0$ and 1 correspond to the horizontal 
       and vertical directions, respectively. The wavelength is $\lambda=0.948\,\mu$m.
       \label{F9}}
\end{figure}

The angular distribution of the emergent radiation, known as limb darkening, is affected by refraction.
Limb darkening is shown in the first panel of Fig. \ref{F8} since at $\tau=0$, $P'_\nu(\mu)=P_\nu(\mu)=I_\nu(\mu)$.
In the non-refractive case, we have $I_{\nu}(\tau_{\nu}=0,\mu=0)/I_{\nu}(0,1)=0.423$ for $\lambda=0.948\,\mu$m.
When refraction is introduced, the limb darkening is much weaker and $I_{\nu}(0,0)/I_{\nu}(0,1)=0.947$.   
In the presence of refraction, ray paths that emerge with $\mu \sim 0$ have been strongly deflected. 
The vertical optical depth $\tau_{\nu}$ for a ray path exiting with $\mu=0$ is related to $\sigma_{\nu}$ by
\begin{equation} 
    \tau_\nu=\int_0^{\sigma_\nu}{\sqrt{1-1/n_\nu^2(\sigma'_\nu)}\,d\sigma'_\nu}.
\end{equation}
For $n_{\nu}(\tau_{\nu})$ shown in Fig. \ref{F1} and $\sigma_\nu=1$, we get $\tau_\nu=0.56$. 
This means that due to refraction, a horizontal viewing angle allows us to see as deep inside the 
atmosphere as under angle $\mu=0.56$ in the non-refractive case. We can indeed see on Fig. \ref{F8} 
that $I'_\nu(0,0)_{\rm ref} = I'_\nu(0,0.56)_{\rm non\ ref}$.
\begin{figure}[t]
\centering
\epsscale{0.6}
\plotone{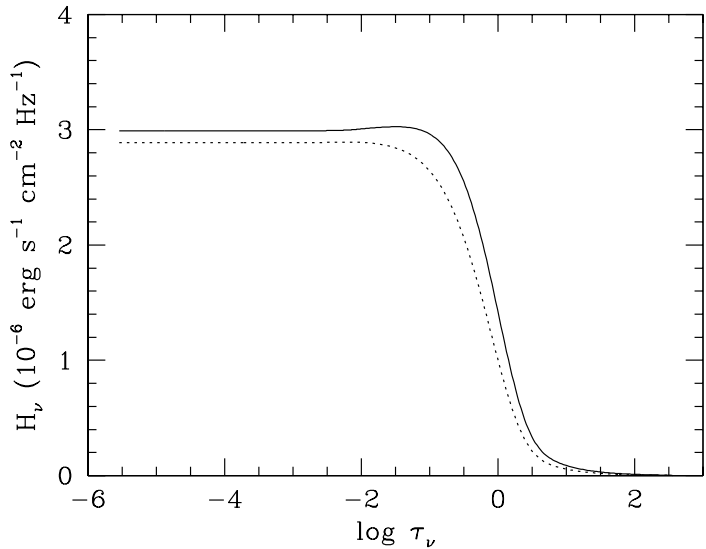}
\figcaption[P6.ps]{Radiative flux for refractive (solid line) and non-refractive 
      (dotted line) cases as a function of vertical optical 
       depth in the nominal white dwarf atmosphere model for $\lambda=0.948\,\mu$m.
       The atmospheric structure is the same for both calculations.
      \label{F8}}
\end{figure}
 
Despite the significant change in the specific intensity shown in Fig. \ref{F9}, 
the radiative fluxes are nearly identical throughout the atmosphere (Fig. \ref{F8}). 
This happens because in the absence of absorption and emission, refraction is a geometric effect that
does not affect the value of the integral (\ref{321}).
By symmetry, the reflected ray paths do not contribute to the flux ($I'_{\nu}(+\mu)=I'_{\nu}(-\mu)$) and
Snell's law implies that $n^{2}_{\nu}\mu d\mu$ is constant.    
For reflected ray paths in an actual atmosphere, $I'_\nu(\tau_\nu,\mu>0)>I'_\nu(\tau_\nu,\mu<0)$, as emission 
is usually larger deeper inside the atmosphere and absorption processes are present. This explains why
the radiative flux $H_\nu$ is slightly larger in the refractive case. 
For $\tau_\nu>0.1$, the radiative flux decreases rapidly due to 
the presence of a convective zone, a characteristic of all cool white dwarf atmospheres models.
\begin{figure}[t]
\centering
\epsscale{0.9}
\plotone{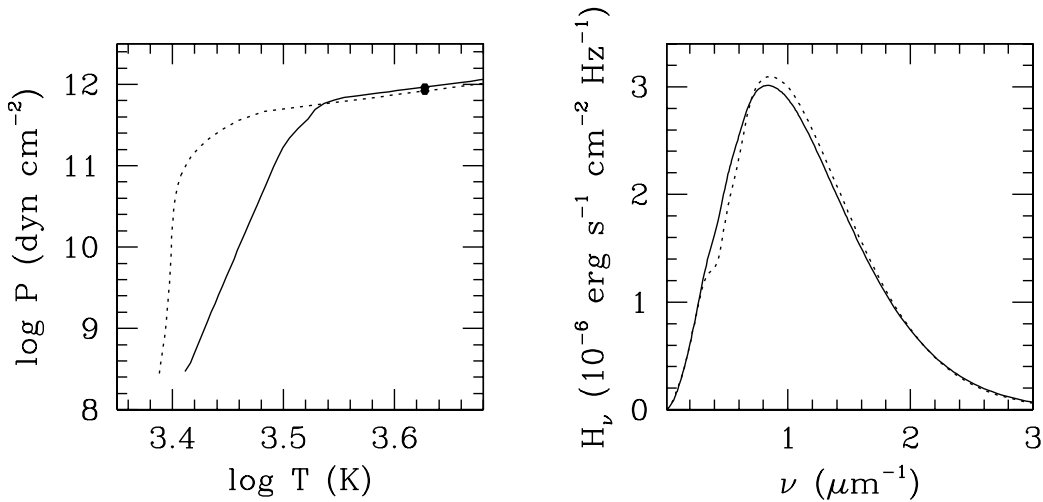}
\figcaption[P7.ps]{Pressure-temperature structure (left panel) and synthetic spectrum (righ panel)
       for the nominal white dwarf atmosphere model parameters 
       with (solid line) and without (dotted line) dispersive effects.
       Both models are computed by imposing flux conservation and hydrostatic equilibrium.
       The filled circle indicates the level where the Rosseland mean optical depth $\tau_{\sss R}=1$.
       \label{F10}}
\end{figure}
  
So far, our discussion of the effects of refraction on the radiation field was based on a fixed 
(non-refractive) atmospheric pressure-temperature structure.  A full atmosphere calculation consists 
of finding the structure that satisfy the 
equations of 1) radiative transfer, 2) flux constancy in convection zones and
radiative equilibrium in radiative zones,
and 3) hydrostatic equilibrium.\footnote{The atmosphere model uses a realistic helium equation 
of state which predicts a very low degree of ionization.  Electron conduction in the atmosphere is
therefore negligible (Bergeron et al. 1995), in contrast with the models of Kapranidis (1983).}
Figure \ref{F10} compares such a self-consistent calculation for 
both the non-refractive (nominal structure) and the refractive cases. As expected, internal reflection 
leads to a hotter structure near the surface.
The spectra resulting from these self-consistent structures are shown in Fig. \ref{F10}.
The $0.4\,\mu$m$^{-1}$ feature seen in the non-refractive case is caused by collision-induced 
absorption (CIA) by the small amount 
of H$_2$ molecules present in this model ($n({\rm H}_2)/n({\rm He})\sim4\times 10^{-7}$). In 
the refractive case, the hotter structure
results in the dissociation of $\rm H_{2}$ and reduces the CIA absorption. 
The effects of refraction become much larger in models with lower effective temperatures or higher 
gravities and generally decreases as the hydrogen abundance increases.
A detailed study of refractive effects in very cool white dwarfs atmospheres and spectra 
will be the subject of a subsequent publication.

\section{Conclusions}

Having realized that in the dense atmospheres of very cool white dwarf stars -- especially those 
with a composition dominated by helium -- the index of refraction departs significantly from 
unity, we have solved the problem of radiative transfer in plane parallel, LTE,
static stellar atmospheres in the presence of refraction for arbitrary vertical 
variations of the opacity, temperature, and index of refraction. We show how to 
modify the Feautrier solution and the $\Lambda$-iteration method to solve the 
equation of radiative transfer in the presence of refraction. The resulting methods are general and 
can be applied to any one-dimensional problem.
For cool white dwarfs, the most important effect of refraction on the propagation of 
radiation in a dispersive medium is total internal reflection.  This increases the mean 
intensity in the optically thin region near the surface, where the dispersive effects are strongest.
If radiative equilibrium is imposed, this results in higher surface temperatures than in the non-refractive case. 
We also find that limb darkening is dramatically reduced by refraction.
This first solution of the radiative transfer equation with 
dispersive effects in stellar atmospheres suggest that refraction is an 
important effect that must be considered when modeling the atmospheres 
of very cool white dwarfs. 

The authors wish to thank D. Mihalas and L.H. Auer for useful discussions. P. Kowalski is grateful to the
Los Alamos National Laboratory for its hospitality and support. The research 
presented here was supported by NSF grant AST97-31438, NASA grant NAG5-8906, and by the United States 
Department of Energy under contract W-7405-ENG-36.

\clearpage
\Large

\end{document}